\documentclass[aps,twocolumn,amssymb]{revtex4}
\usepackage{graphicx}
\usepackage{longtable}
\usepackage{epsfig}
\usepackage{dcolumn}
\usepackage{bm}

\begin{document}

\title{Comparative study of the electronic structures of Fe$_3$O$_4$ and Fe$_2$SiO$_4$}

\author{      Przemys\l{}aw Piekarz }
\affiliation{ Institute of Nuclear Physics, Polish Academy of Sciences, 
              Radzikowskiego 152, 31-342 Krak\'{o}w, Poland }

\author{      Andrzej M. Ole\'{s} }
\affiliation{ Institute of Nuclear Physics, Polish Academy of Sciences, 
              Radzikowskiego 152, 31-342 Krak\'{o}w, Poland }

\author{      Krzysztof Parlinski }
\affiliation{ Institute of Nuclear Physics, Polish Academy of Sciences, 
              Radzikowskiego 152, 31-342 Krak\'{o}w, Poland }
	  
\begin{abstract}
The electronic properties of two spinels Fe$_3$O$_4$ and Fe$_2$SiO$_4$ are studied
by the density functional theory. The local Coulomb repulsion $U$ and the Hund's
exchange $J$ between the $3d$ electrons on iron are included.
For $U=0$, both spinels are half-metals, with the minority $t_{2g}$
states at the Fermi level. Magnetite remains a metal in a cubic phase even at large 
values of $U$. The metal-insulator transition is induced by the $X_3$ phonon,
which lowers the total energy and stabilizes the charge-orbital ordering.
Fe$_2$SiO$_4$ transforms to a Mott insulating state for $U>2$ eV with 
a gap $\Delta_g\sim U$. The antiferromagnetic interactions induce the tetragonal
distortion, which releases the geometrical frustration and stabilizes the 
long-range order. The differences of electronic structures in the high-symmetry 
cubic phases and the distorted low-symmetry phases of both spinels are discussed.  
\end{abstract}

\maketitle 

\section{Introduction}

In transition-metal spinels, magnetic, charge, and orbital 
interactions are strongly affected by the crystal geometry.
A special character of spinels is associated with
the arrangement of cations in the octahedrally coordinated sites
forming a pyrochlore lattice. Consisting of the corner-sharing
tetrahedra, this crystal structure causes the geometrical frustration
of magnetic interactions and makes the electronic
ground state highly degenerate \cite{Anderson}. 
In such cases, the system may remain in a disordered state 
like a quantum spin liquid or 
spin ice even at low temperatures \cite{Spin_Ice}.
It makes also the electronic state very susceptible to lattice 
distortion, which lifts the degeneracy and drives the 
structural phase transition. 

\begin{figure}[b]
\includegraphics[width=8cm]{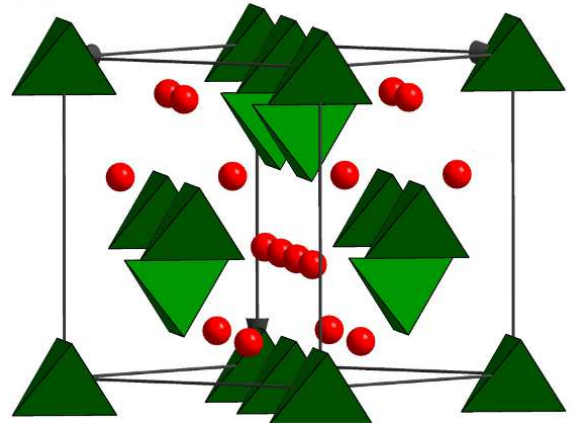}
\caption{The spinel structure: the octahedral B sites are represented
by red balls and the AO$_4$ units are plotted as green tetrahedra.}
\label{fig:fig1}
\end{figure}

Spinels have a general formula AB$_2$O$_4$ and at room temperature
crystallize in the cubic $Fd\bar{3}m$ ($O^7_h$) symmetry (Fig. \ref{fig:fig1}).
In a normal spinel, the A and B cations occupy the tetrahedral and octahedral
positions, respectively. 
The B sites create the atomic chains along the $[110]$ and $[1\bar{1}0]$
directions.
In the most common cases, the valence state for the A site is 2+ and 
for the B site 3+ ($e.g.$ in MgAl$_2$O$_4$).
The 4+ valence state for the A site exists only in the high-pressure 
forms of olivines (Fe$_2$SiO$_4$, Co$_2$SiO$_4$).
In inverse spinels, the A atoms occupy half of the octahedral
sites, while the B ones are shared equally by the octahedral
and tetrahedral sites. Fe$_3$O$_4$ is classified as the inverse spinel
due to mixed valency of iron ions in the octahedral sites (2+/3+).

Depending on the type of cations in the octahedral sites,
transition-metal spinels exhibit various types of ordered
phases at low temperatures \cite{Radaelli}. 
One observes antiferromagnetic (AF) phases (MgV$_2$O$_4$) \cite{Niziol}, 
charge ordering (CuIr$_2$S$_4$) 
\cite{CO}, metal-insulator transitions (MgTi$_2$O$_4$) \cite{MI}, 
or spin-Peierls transitions (ZnCr$_2$O$_4$) \cite{SP}. 
Crystal distortions in these phase transitions are associated 
with different electron-lattice couplings.
In ZnMn$_2$O$_4$, the symmetry reduction to the tetragonal space group $I4_1/amd$
is driven by the Jahn-Teller effect.
The tetragonal distortion in ZnCr$_2$O$_4$ observed at $T_N=12.5$ K 
induces the spin gap in magnetic excitations similarly as in 
the one-dimensional spin-Peierls systems.
The metal-insulator transition in MgTi$_2$O$_4$ is associated
with the spin-singlet formation and dimerization of the atomic bonds \cite{Singlet}. 

The Verwey transition in Fe$_3$O$_4$ has a very unique character
and its microscopic origin has been intensively studied for
more than 60 years. 
Verwey proposed that a discontinuous drop of the electrical 
conductivity at $T_V=122$ K is associated with the ordering
of the Fe$^{3+}$ and Fe$^{2+}$ ions in the octahedral sites \cite{Verwey}.
According to this scenario, the conductivity increases above $T_V$ due 
to higher mobility of electrons fluctuating between the Fe sites.
At $T_V$, electron movement freezes out and the charge ordering
stabilized by the electrostatic forces develops. However, this simple 
picture was not confirmed by more recent studies.
Firstly, Anderson demonstrated that in the spinel geometry the energy 
difference between a short-range and long-range order is very small 
and therefore typical order-disorder mechanism is therefore unlikely 
\cite{Anderson}. Secondly, the crystallographic studies revealed the 
monoclinic deformation below $T_V$, which is inconsistent with the 
Verwey model \cite{Iizumi}. The diffraction measurements support only 
the fractional charge ordering with small differences of charges 
\cite{Attfield,Nazarenko}.
Finally, the theoretical studies demonstrated the importance of local 
Coulomb interactions in the $3d$ states on iron, which induces the
charge-orbital ordering in the monoclinic phase \cite{Leonov}.
In our studies, we developed the mechanism of the Verwey transition based on the
interplay between the electron correlations and the electron-phonon
interaction \cite{PPO1,PPO2,PPO3}.

Spinels containg Si atoms belong to a separate 
group of minerals stable only under high pressure, as realized in the 
Earth upper mantle.
Fe$_2$SiO$_4$, which is the end-member of geologically import 
group of olivines (Mg,Fe)$_2$SiO$_4$, can be obtained from magnetite 
by replacing the Fe atom at the A site by the Si atom.
The experimental information on the electronic properties
of this mineral is very limited. The heat capacity measurement
revealed the peak at $T_N=11.8$ K, most probably associated
with the paramagnetic-antiferromagnetic transition \cite{HC}.
Above $T_N$, a paramagnetic state is consistent with
the M\"{o}ssbauer measurements \cite{Mossbauer}.
In the previous study, we have analyzed the electronic
structure of Fe$_2$SiO$_4$ using the density functional theory 
(DFT) \cite{Mariana}.
We found that the insulating state results from the local Hubbard
interaction $U$ included within the generalized gradient approximation 
within the so-called GGA+$U$ approach. In addition, we revealed that 
the AF interactions
induces the tetragonal distortion, removing the geometrical frustration.
In this paper, we compare the electronic properties
of Fe$_3$O$_4$ and Fe$_2$SiO$_4$, analyzing in details the
changes induced by the Hubbard interaction $U$ and crystal distortions.
For Fe$_2$SiO$_4$, we investigate the effect of high pressure
on the electronic structure.

The paper is organized as follows. The details of the calculation method 
are presented in Sec. \ref{sec:cal}. Electronic structures of Fe$_3$O$_4$ 
and Fe$_2$SiO$_4$ are reported and analyzed separately for the high 
symmetry phase (Sec. \ref{sec:high}) and low symmetry phase (Sec. 
\ref{sec:low}). Discussion of the results and a short summary are
presented in \ref{sec:summa}.

\section{Calculation method}
\label{sec:cal}

The electronic and crystal structures of both spinels were optimized 
within the GGA+$U$ method using the VASP code \cite{Vasp}.
The wave functions were obtained by the full-potential 
projector augmented-wave method \cite{PAW}.
The basis included the following valence electron configurations:
Fe:$3d^64s^2$, Si:$3s^23p^2$, and O:$2s^22p^4$.
To compare the effects induced by the local electron interactions
in the $3d$ states on iron, the same values of the Coulomb 
and Hund's exchange parameters were chosen: $U=4$ eV and $J=0.8$ eV. 
The repulsion energy $U$ was taken from the constrained
DFT calculations \cite{Band} and $J$ was obtained from 
the atomic values of the Racah parameters for Fe$^{2+}$ ions \cite{Zaa90}: 
$B=0.131$ eV and $C=0.484$ eV. 
It is reasonable to assume that the Coulomb interaction parameters $\{U,J\}$
are the same in both Fe$_3$O$_4$ and Fe$_2$SiO$_4$ systems in their 
insulating phases, as one expects that the screening of the Coulomb 
interaction should be rather similar. However, we cannot exclude that 
the Coulomb parameter $U$ changes in the magnetite at the Verwey 
transition, when the metallic state is transformed into an insulating 
one (this effect was not considered in the theory until now).

The calculations for the cubic symmetry $Fd\bar{3}m$ were performed
in the $a\times a\times a$ supercell and for
the monoclinic phase $P2/c$ of magnetite in the
$a/\sqrt{2}\times a/\sqrt{2}\times 2a$ cell, both with 56 atoms.
The electronic structures were obtained for the fully
relaxed crystals. For Fe$_2$SiO$_4$, the results were obtained 
at two hydrostatic pressures $p=0$ and $p=20$ GPa.
The latter value corresponds to a stable spinel phase
existing only under high-pressure conditions. 

\section{Electronic structures}
\label{sec:elec}

\subsection{High-symmetry phases}
\label{sec:high}

In the case of the iron-based spinels, the electronic structures can be
well understood starting from the electron configurations 
of the Fe$^{2+}$ and Fe$^{3+}$ ions.
In Fe$_2$SiO$_4$, the iron atoms have a uniform electron distribution
corresponding to the $2+$ charge valency. 
The octahedral sites in magnetite have the mixed valence configuration
with an average charge $2.5+$.
Due to the Hunds's exchange both ionic configurations correspond
to the high spin states Fe$^{2+}$: t$^3_{2g\uparrow}$t$_{2g\downarrow}$
e$^2_{g\uparrow}$ (S=2) and Fe$^{3+}$: t$^3_{2g\uparrow}$e$^2_{g\uparrow}$
(S=5/2).
The splitting of the $3d$ states into the $t_{2g}$ and $e_g$ orbitals
is induced by the crystal field, which is much larger at the octahedral sites 
than at the tetrahedral ones.
The exchange coupling causes the splitting of the majority up-spin
and minority down-spin states and shifts one type of the states with respect to 
the other. Consequently only the down-spin states occupy the energy levels 
close the Fermi energy ($E_F$) \cite{Band}.

\begin{figure}[t]
\includegraphics[width=8cm]{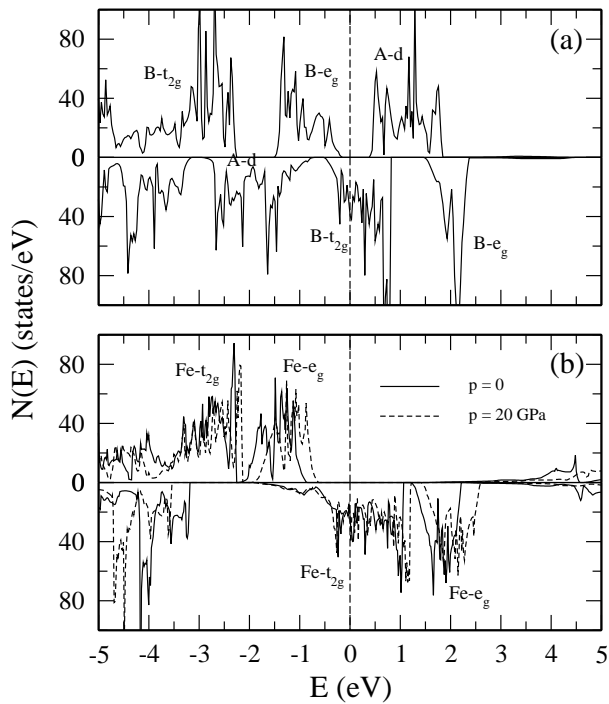}
\caption{Electronic structure of (a) Fe$_3$O$_4$ at $p=0$ pressure 
and (b) Fe$_2$SiO$_4$ obtained in the cubic $Fd\bar{3}m$ symmetry for two pressure values, 
$p=0$ (solid lines) and $p=26$ GPa (dashed lines), and for $U=J=0$.
}
\label{fig:fig2}
\end{figure}

In Fig. \ref{fig:fig2}, we compare the electronic density of states (DOS)
in these compounds for $U=J=0$. If the local electron interactions are
neglected, the grounds states of both spinels are half-metallic with 
the minority down-spin $t_{2g}$ states at $E_F$. These states come 
only from the Fe atoms at the octahedral sites. Neither the $3d$ states 
on the Fe(A) atoms in magnetite nor the $s$ and $p$ states participate 
in the electron transport.
Similarly to the Fe(B) states, the up-spin and down-spin states
at the Fe(A) sites are splitted by the exchange coupling.
All the down-spin states are occupied, while the up-spin states
are empty, so the magnetic moments on the A and B sites have
the opposite directions generating the ferrimagnetic order in magnetite.
The half-metallic ground state of magnetite agrees with the properties of 
the high-symmetry phase indicating that the local Coulomb interactions
do not play important role at high temperatures.

Without the Hubbard interaction, the metallic ground state of
Fe$_2$SiO$_4$ has the ferromagnetic (FM) order with the magnetic moment 
on iron equal $3.57 \mu_B$.
It remains a metal even at high pressures, showing a disagreement
with the insulating behavior of this material.
As shown in Fig. 2(b), the main effect of pressure is to increase the
bandwidth due to larger overlap of the $3d$ orbitals. 
The splitting of the $t_{2g}$ and $e_g$ states also increases because of 
the stronger crystal field.

In the next step, we compare the changes in the DOSs
induced by the on-site interactions $U$ and $J$, while 
keeping the cubic symmetry. We allow only for the relaxation of the 
lattice constant and the internal parameter $u$. The lattice constants 
of both crystals increase slightly (less than $1\%$) because of the 
larger Coulomb repulsion between electrons \cite{PPO2,Mariana}. 
In magnetite, 
the electronic state does not change much comparing to the uncorrelated 
case, Fig. 3(a). There is an additional shift of the majority up-spin 
Fe(B) and down-spin Fe(A) states to lower energies due to larger 
exchange coupling. It increases the magnitudes of magnetic moments on 
the Fe(A) sites from 3.40 to 3.97 $\mu_B$ and on the Fe(B)
sites from 3.52 and 3.87 $\mu_B$, improving the agreement with
the experiment. The ground state remains metallic even at presence
of strong on-site interaction $U$.

\begin{figure}[t]
\includegraphics[width=8cm]{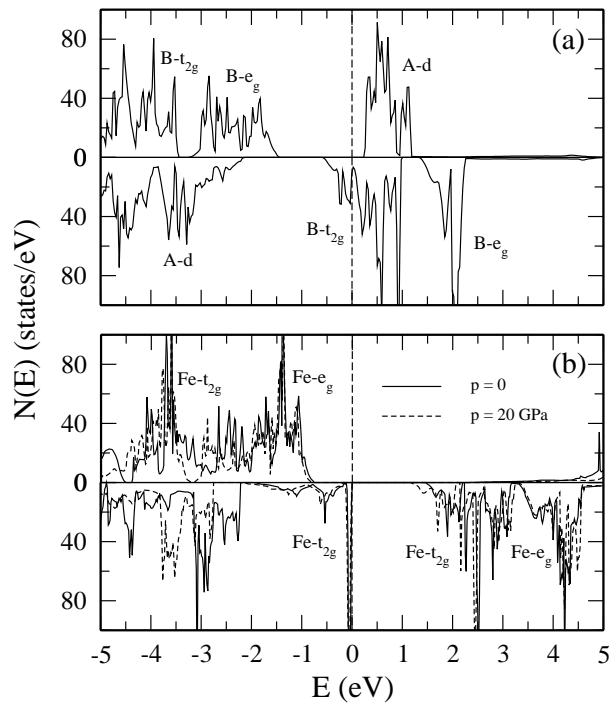}
\caption{The same as in Fig. 1 for $U=4$ eV and $J=0.8$ eV.}
\label{fig:fig3}
\end{figure}

In Fe$_2$SiO$_4$, there is a larger electron occupation of the $t_{2g}$ 
states on the Fe(B) atoms (one electron per one iron site),
so the on-site Hubbard interaction $U$ is much more effective
than in magnetite. As shown in Fig. 3(b), the interaction $U$ has a strong
impact on the electronic DOS, inducing the insulating gap at $E_F$.
The gap with the magnitude $1.5$ eV separates the occupied down-spin
$t_{2g}$ states from the empty ones. The insulating state 
is generated by the inter-orbital Hubbard interaction with the effective 
energy $U_{eff}=U-J$ (note that $U$ is here an average value of local
Coulomb interaction for different electron pairs, as usually assumed in 
the local density approximation supplemented by local electron-electron 
interactions (LDA+$U$) \cite{Leonov,Lic95}),
so the magnitute of the gap can be estimated from the formula 
\begin{equation}
\Delta_g=U_{eff}-\frac{W}{2},
\label{eq:Mott}
\end{equation}
where $W$ is the width of the $t_{2g}$ band. 
For $W\sim 3$ eV, we get $\Delta_g\sim 1.7$ eV, which corresponds 
very well with the obtained gap.
At high pressure, the insulating gap decreases to 1.2 eV, 
and this effect can be easily understood from Eq. (\ref{eq:Mott}) ---
under increasing pressure the band width $W$ increases, thus reducing 
the value of the gap.

The results obtained for a cubic phase with a realistic value of 
$U$ reveal a fundamental difference between these two spinel 
structures. On the one hand,
Fe$_2$SiO$_4$ behaves like a Mott insulator with the gap 
$\Delta_g\sim U$,
and the local Hubbard interaction is crucial to reproduce
the electronic properties of this material. On the other hand,
the on-site interaction $U$ alone does not induce the insulating 
state in magnetite, and in the cubic symmetry this material 
remains in a metallic state. This may be concluded from
the behavior at high temperatures, in particular
from a relatively high conductivity.
This theoretical result indicates that Fe$_3$O$_4$ is not a Mott insulator,
and its non-metallic state at low-temperatures has a different origin.

\subsection{Low-symmetry phases}
\label{sec:low}

\begin{figure}[b]
\includegraphics[width=8.4cm]{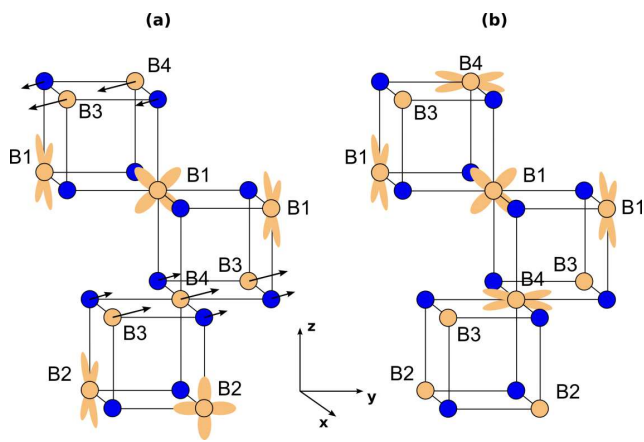}
\caption{Orbital ordering induced by 
(a) the $X_3$ mode (the black arrows), and 
(b) the monoclinic symmetry $P2/c$.
Nonequivalent Fe positions are labeled B1, B2, B3, and B4.
}
\label{fig:fig4}
\end{figure}

The phase transitions observed in the discussed materials at low 
temperatures have fundamentally different nature. The Verwey transition 
involves both the charge and orbital degrees of freedom, and their 
ordering below $T_V$ is stabilized by the monoclinic distortion 
\cite{Leonov}.
Using the group theory, we found that the low-symmetry phase results
from the condensation of two phonons (order parameters) with the 
symmetries $\Delta_5$ and $X_3$ \cite{PPO1}. The strength of the 
electron-phonon 
coupling strongly depends on the Hubbard interaction $U$, which 
induces the antiferro-orbital ordering and splitting of the $t_{2g}$ 
states at the Fermi level. Without the lattice distortion, 
the electronic state with orbital polarization
is highly degenerate showing a short-range order \cite{Pinto}.
We found that the charge-orbital ordering is stabilized by the 
coupling between the $3d$ electrons and the $X_3$ phonon \cite{PPO2}.
This coupling lowers the total energy of the crystal and induces
the metal-insulator transition.

The orbital order of the $t_{2g}$ states with the atom displacements 
corresponding to the $X_3$ mode is shown in Fig. 4(a). The positions 
of atoms were generated from the polarization vectors using the PHONON
software \cite{Phonon}. Iron atoms occupy 4 different positions in
the strucure labeled as B$_1$, B$_2$, B$_3$, and B$_4$.
Phonons at the $X$ point are doubly degenerate, so we have chosen
only one phonon branch.
In the presented case, the displacements of the B$_3$ and B$_4$ atoms 
together with oxygens located in the same plane induce the orbital 
ordering on the B$_1$ and B$_2$ atoms in the nearest neighbor planes.
The second order parameter $\Delta_5$ also couples strongly to 
electrons but it does not induce the gap opening.
It generates the crystal distortion which doubles the unit cell
in the $c$ direction.
These results demonstrate that the cubic phase is not a real ground state
of magnetite, and the structural transition to a phase with broken symmetry
must occur when temperature is lowered. 

\begin{figure}[t]
\includegraphics[width=8cm]{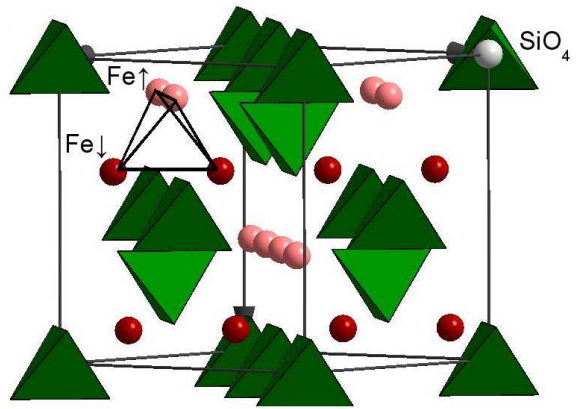}
\caption{The AF ordering in Fe$_2$SiO$_4$ --- light/dark spheres 
indicate positions of $\uparrow$ and $\downarrow$ magnetic moments 
on Fe atoms.}
\label{fig:fig5}
\end{figure}

In contrast, the phase transition occurs in Fe$_2$SiO$_4$ at 
much lower temperature $T_N=11.8$ K, and is driven by the magnetic 
interactions. We have considered two magnetic configurations with the 
FM and AF order. We assumed that the magnetic moments in the AF 
configuration are aligned parallely in the (001) planes, and have 
opposite (alterneting) directions in the subsequent planes along the 
$c$ axis (see Fig. 
\ref{fig:fig5}). Instead of the $c$ axis one can choose either $a$ or 
$b$ direction, so the AF order is triply degenerate in the cubic 
symmetry. We found that for $U>2$ eV, there is a change of the 
ground state from the FM to AF order \cite{Mariana}. This change is 
driven by the metal-insulator transition at the same value of $U$.
Therefore, the electronic DOS with the FM order presented in Fig. 3(b) 
does not correspond to the ground state realized at $U=4$ eV. We 
revealed that the AF configuration induces the tetragonal distortion 
($a>c$), breaking the cubic symmetry and lowering the total energy.
This distortion depends on the Hubbard interaction $U$ and for $U=4$ eV 
one finds $c=0.985a$. The contraction of the crystal along the $c$ 
axis enhances the superexchange interaction along the AF bonds,
while the elongation in the $ab$ plane weakens the one along FM 
bonds.

The electronic DOSs in the low symmetry phases of both spinels were 
obtained by full relaxation of the crystal structures without imposing 
the cubic symmetry constrains. Fe$_3$O$_4$ has the monoclinic $P2/c$ 
symmetry with the doubled lattice constant in the $c$ direction. In 
this symmetry there are two non-equivalent tetrahedral positions A$_1$ 
and $A_2$ (not shown), and four octahedral sites B$_1$, B$_2$, B$_3$, 
and B$_4$, shown in Fig. 4(b). 
The lattice parameters and atomic positions were analyzed in detail in 
Ref. \cite{PPO2}. Fe$_2$SiO$_4$ has the tetragonal $I4_1/amd$ space 
group. 

\begin{figure}[t]
\includegraphics[width=8.4cm]{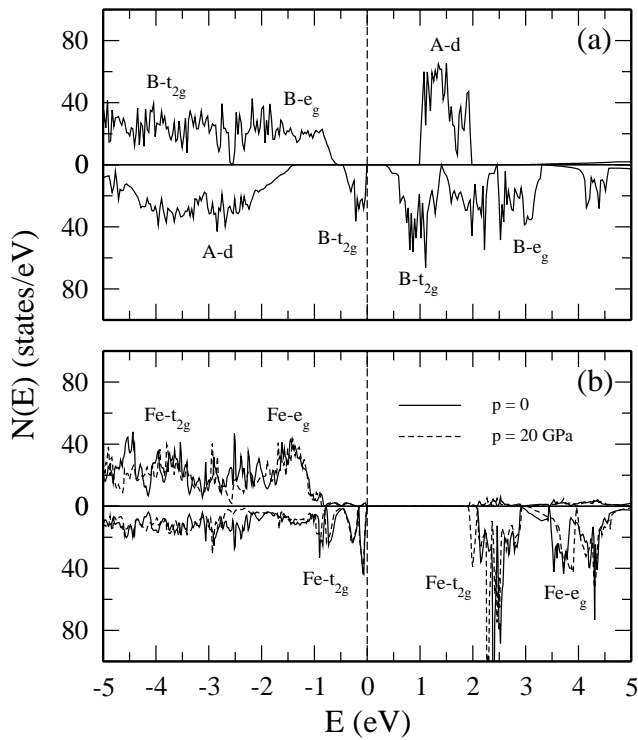}
\caption{
Electronic structures of (a) the $P2/c$ phase of Fe$_3$O$_4$, and
(b) the $I4_1/amd$ phase of Fe$_2$SiO$_4$.
Parameters: $U=4$ eV and $J=0.8$ eV.}
\label{fig:fig6}
\end{figure}

The electronic structures for the above symmetries are presented in 
Fig. \ref{fig:fig6}. 
The insulating gaps in the minority $t_{2g}$ states now exist in both
systems. Comparing to the FM phase of Fe$_2$SiO$_4$, the gap in the AF 
phase is larger (2 eV). Under pressure, the magnitude of the gap 
decreases as in the FM case. The gap $\Delta_g=0.33$ eV obtained in
magnetite agrees quite well with the experimental value 0.14 eV 
\cite{Gap} and is much smaller than the gap in Fe$_2$SiO$_4$. The gap 
separates the occupied states at the B$_1$ and B$_4$ sites from the 
empty ones at the B$_2$ and B$_3$ sites. This explains the origin of the 
fractional charge ordering in the monoclinic phase. The orbital ordering 
arises due to the preferential occupation of the $t_{2g}$ states. 
A schematic plot of the orbital ordering is presented in Fig. 4(b).
In the B$_1$ chains, the electrons occupy the $d_{xz}$ and $d_{yz}$ 
orbitals creating the antiferro-orbital (alternating orbital) order.
At the B$_4$ sites only the $d_{xy}$ orbitals are occupied by electrons.
 
\section{Discussion}
\label{sec:summa}

The presented results reveal some interesting aspects of two iron-based
spinels. As a common feature, the electronic properties of both 
compounds are strongly affected by the on-site Hubbard interaction $U$. 
The interplay of this local Coulomb interaction with a lattice
distortion is the driving force of the observed phase transitions, 
which lead in both spinels to a new ordered phase with a broken cubic 
symmetry.

In magnetite, this interaction does not lead directly to the 
insulating phase as in the Mott-Hubbard mechanism. It enhances
a natural tendency of the degenerated $t_{2g}$ states to create
an orbital ordering. Such state becomes very sensitive to lattice
distortion, which by splitting the $t_{2g}$ band opens the insulating
gap. This mechanism is similar to the Jahn-Teller effect,
however, the lattice distortion in magnetite is much more complex
than in simple perovskites, see e.g. Ref. \cite{Ree06}. The structural 
transition is here induced by two order parameters, which couple to each 
other and generate the monoclinic phase \cite{PPO1}.

Long time ago, Anderson predicted the existence of the short-range 
order in magnetite on the basis of the simple electrostatic model 
\cite{Anderson}. There are many signs of such state above the Verwey 
transition. One of them is the insulating gap that does not close 
completely at high temperatures \cite{Park}. Recently, the resonant 
x-ray scattering studies revealed that charge-orbital ordering starts 
to develop about 10 K above $T_V$ \cite{Lorenzo}.
The charge-orbital fluctuations couple to phonons producing
the critical diffuse scattering. The neutron studies found the 
symmetries of lattice distortions associated with these fluctuations. 
They are the precursors of the order parameters $\Delta_5$ \cite{Fujii} 
and $X_3$ \cite{Siratori}.

We found that the geologically important mineral Fe$_2$SiO$_4$-spinel
is a Mott insulator. The on-site Hubbard interaction $U$ in the $3d$ 
states on iron generates the insulating state with the AF order. We
suggest that these AF interactions drive the phase transition observed 
at $T_N=11.8$ K. Furthermore, we discovered that the AF order induces 
the tetragonal distortion reducing the cubic $Fd\bar{3}m$ symmetry to 
its subgroup $I4_1/amd$. This distortion releases the geometrical
frustration of the magnetic interaction enabling for the long-range 
order. In Ref. \cite{Mariana}, we have estimated the exchange constant 
$J$ and the Curie-Weiss temperature $\theta_{\rm CW}$ using the 
classical Heisenberg model. The obtained large values of the exchange
constant $J=1.7$ meV and the Curie-Weiss temperature $\theta_{CW}=340$ 
K are in sharp constrast with the low value of $T_N=11.8$ K and 
indicate that magnetic interactions are strongly 
frustrated and are expected to generate short-range order above $T_N$. 

So far, there are no experimental indications that the AF order in 
its simplest form (as in Fig. \ref{fig:fig5}) exists below $T_N$,
and its nature is more subtle. 
One possibility is that the magnetic order is non-colinear,
with complicated spin orientations. In such a case, the low-symmetry 
structure could be different from the tetragonal. It is also possible 
that below $T_N$ the magnetic order has only a short-range character 
(similar to spin ice), or the long-range order coexists with the 
short-range order like in the ZnFe$_2$O$_4$ spinel \cite{Schiessl}. 

In conclusion, we suggest that while the origin of the Verwey 
transition in Fe$_3$O$_4$ could be understood, the nature of the 
magnetic order in Fe$_2$SiO$_4$ below $T_N$ poses an interesting 
experimental problem. We hope that future experiments will either 
confirm the predictions of the present studies, or indicate necessary 
ingredients of a more complete theoretical treatment of this problem.

\acknowledgments

The authors thank all collaborators Mariana Derzsi, Pawe{\l} Jochym,
Jan {\L}a\.{z}ewski, and Ma{\l}gorzata Sternik for valuable discussions.
This work was supported in part by Marie Curie Research Training
Network under Contract No. MRTN-CT-2006-035957 (c2c).
A.M. Ole\'s acknowledges support by the Foundation for Polish Science 
(FNP) and by the Polish Ministry of Science and Higher Education under
Project No.~N202 068 32/1481.

\end{document}